\documentclass[12pt]{article}

\catcode`\@=11
\global\arraycolsep=2pt
\usepackage{amsbsy,amssymb,latexsym,amsfonts,amsmath}
\usepackage{graphicx,color}

\begin{document}
\begin{flushright}
\parbox{4.2cm}
{RUP-20-3}
\end{flushright}

\vspace*{0.7cm}

\begin{center}
{ \Large  CP-violating super Weyl anomaly}
\vspace*{1.5cm}\\
{Koichiro Nakagawa and Yu Nakayama}
\end{center}
\vspace*{1.0cm}
\begin{center}

Department of Physics, Rikkyo University, Toshima, Tokyo 171-8501, Japan

\vspace{3.8cm}
\end{center}

\begin{abstract}
In CP-violating conformal field theories in four dimensions, the Pontryagin density can appear in the Weyl anomaly. The Pontryagin density in the Weyl anomaly is consistent, but it has a peculiar feature that the parent three-point function of the energy-momentum tensor can violate CP only (semi-)locally. In this paper, we study the supersymmetric completion of the Pontryagin density in the Weyl anomaly, where the central charge $c$ effectively becomes a complex number. The supersymmetry suggests that it accompanies the graviphoton $\theta$ term associated with the R-symmetry gauging in the Weyl anomaly. It also accompanies new CP-violating terms in the R-current anomaly. While there are no conclusive perturbative examples of CP-violating super Weyl anomaly, we construct explicit supersymmetric dilaton effective action which generates these anomalies.
\end{abstract}

\thispagestyle{empty} 

\setcounter{page}{0}

\newpage


%-----------------------------------------

%\pacs{}
%\keywords{Suggested keywords}%Use shokeys class option if keyword
%\maketitle

%%%%%%%%%%%%%%%%%%%%%%%%%%%%%%%%%%%%%%%%%%%%%%%%%%%%
\section{Introduction}
Charge-Parity (CP) symmetry and its breaking play fundamental roles in our understanding of our existence. It is the violation of the CP symmetry in the electro-weak sector that enables us to generate non-zero baryon numbers while we mysteriously observe the highly suppressed CP violation (beyond what we can explain by the anthropic principle) in the strong interaction sector of the standard model of particle physics. Within the Lagrangian description of quantum field theories, one can quantify the violation of the CP symmetry by examining the detailed structure of coupling constants: Kobayashi-Masukawa matrix and $\theta$-term are such quantities in the electro-weak and strong interaction sectors in the standard model.

It is more desirable to establish a quantification of the violation of the CP symmetry beyond the Lagrangian description, which can be used, for instance, in strongly coupled conformal field theories. In conformal field theories, we often use conformal data that are related to anomalies to quantify non-perturbative characteristics of the theories. For example, the number of degrees of freedom, naively counted by a number of fields, is replaced by the central charge that appears in the Weyl anomaly, for which we can rigorously prove that it decreases along the renormalization group flow. Chiral asymmetries in charged objects can be non-perturbatively quantified by the 't-Hooft anomaly coefficients that are invariant under the renormalization group flow.

In \cite{Nakayama:2012gu}, it was pointed out that the Weyl anomaly in four-dimensional conformal field theories may include a term that is present only in CP-violating theories. It is the Pontryagin density. Since the Pontryagin density is CP odd while the energy-momentum tensor is CP even (in CP preserving theories), it can appear only in the CP-violating theories. Therefore one hopes that the Pontryagin density in the Weyl anomaly can measure the violation of the CP symmetry in the strongly coupled conformal field theories.

Most of the anomalies in conformal field theories can be directly computed from the conformal data i.e. two-point functions and three-point functions. In particular, the CP even terms in the Weyl anomaly i.e. the Weyl tensor squared term (central charge $c$) and the Euler density (central charge $a$) can be read from the three-point functions of the energy-momentum tensor \cite{Erdmenger:1996yc}.
The Pontryagin density in the Weyl anomaly, however, has a peculiar feature that the parent three-point function of the energy-momentum tensor can violate CP only (semi-)locally. Such anomalies are called ``impossible anomalies" in \cite{Nakayama:2018dig} due to the non-existence of conformally invariant non-local correlation functions.

%{\it A comment on the debate on one-loop Pontryagin trace anomaly}

Recently there have been active discussions if we may realize the Pontryagin density in the Weyl anomaly in concrete field theory examples. In particular, it was claimed in \cite{Bonora:2014qla} that a free massless Weyl fermion in four-dimensions can generate the Pontryagin density in the Weyl anomaly with an imaginary coefficient (in the Lorentzian signature). If this is the case, it means that the free Weyl fermion breaks the CP symmetry and unitarity anomalously. The computation has been scrutinized under various regularization schemes \cite{Bonora:2014qla}\cite{Bonora:2015nqa}\cite{Bonora:2015odi}\cite{Bonora:2016ida}\cite{Bastianelli:2016nuf}\cite{Bonora:2018obr}\cite{Bastianelli:2018osv}\cite{Frob:2019dgf}\cite{Bastianelli:2019fot}\cite{Bonora:2019dyv}\cite{Bastianelli:2019zrq}, and the discussions seem to remain open.

In this paper, we study the supersymmetric completion of the CP-violating Weyl anomaly in four-dimensional superconformal field theories coupled with background superconformal supergravity. In superconformal field theories, there are various techniques e.g. supersymmetric localization, to compute the partition function, which may eventually help us compute the CP-violating Weyl anomaly in the strongly coupled regime. We will show that with the CP-violating Weyl anomaly, the central charge $c$, together with the Pontryagin density in the Weyl anomaly, is effectively complexified as the gauge coupling constant is complexified with the $\theta$-term in supersymmetric gauge theories. On the other hand, we will see that the central charge $a$ remains real.

As for concrete realizations of the supersymmetric CP-violating Weyl anomaly, we do not offer any conclusive perturbative examples, but we will discuss what would be the supersymmetric extension of the free Weyl fermion with the putative CP  violating Weyl anomaly. We will also show the consistency of the CP-violating super Weyl anomaly explicitly by constructing the supersymmetric dilaton effective action. It can be regarded as a concrete model of the supersymmetric CP-violating Weyl anomaly albeit the Weyl symmetry is spontaneously broken. 

The organization of the paper is as follows. In section 2, we first review the possibility of the Pontryagin density in the Weyl anomaly. We show that the Pontryagin density  satisfies the Wess-Zumino consistency condition, and we can construct the explicit dilaton effective action. Along the way, we also review how the Seeley-DeWitt coefficient of a chiral fermion can include the Pontryagin density. Section 3 is our main contribution. We show that the Wess-Zumino consistency condition for the super Weyl anomaly allows the complexified $c$, which gives the Pontryagin density in the Weyl anomaly in its component form. We also study the structure of the supersymmetric Seeley-DeWitt coefficient. Finally, we explicitly construct the supersymmetric dilaton effective action that gives the supersymmetric CP-violating Weyl anomaly. In section 4, we conclude with some discussions.

\section{CP-violating Weyl anomaly}

\subsection{General structure}
When we put a conformal field theory on a non-trivial four-dimensional manifold, it shows the Weyl anomaly \cite{Capper:1974ed}. Under the infinitesimal Weyl transformation of the metric $\delta g_{mn}  = -2\sigma g_{mn}$, the (effective) action shows the variation of $\delta_{\sigma} S = -\int d^4x \sqrt{g} \sigma T^{m}_{m}$ with the energy-momentum tensor $T_{mn} = \frac{-2}{\sqrt{g}} \frac{\delta{S}}{\delta g^{mn}}$.\footnote{In this paper, unless explicitly stated, we work in Euclidean field theories. The Euclidean action $S$ in our convention is ``negative definite": $S = -\int d^4x \sqrt{g} \partial^m \phi \partial_m \phi$ for a free scalar field so that it appears to be the same in the Lorentzian signature with $(-+++)$ convention. Except that we work in Euclidean field theories, we follow the convention of \cite{Buchbinder:1998qv}\cite{Butter:2013ura}.}  

Since the scaling dimension of the energy-momentum tensor is four in four dimensions, the most general possibility of the Weyl anomaly constructed out of the metric tensor is
\begin{align}
T^{m}_{m} = \tilde{c} \mathrm{Weyl}^2 - \tilde{a}\mathrm{Euler} + \tilde{b} R^2 + \tilde{d}\Box R + \tilde{e} \mathrm{Pontryagin} \ .
\end{align}
Here, 
\begin{align}
\mathrm{Weyl^2} &= C^{mnrs} C_{mnrs}  =  R^{mnrs}R_{mnrs} - 2R^{mn}R_{mn} +\frac{1}{3} R^2
\end{align}
is the Weyl tensor (denoted by $C_{mnrs}$) squared and
\begin{align}
\mathrm{Euler} = R^{mnrs}R_{mnrs}-4R^{mn}R_{mn}+R^2
\end{align}
 is the Euler density. The last term
\begin{align}
\mathrm{Pontryagin} = R^{mnrs}\Tilde{R}_{mnrs} =\frac{1}{2}\epsilon_{mn ab}C^{mnrs}C^{ab}_{\ \ rs}
\end{align}
is the Pontryagin density. It is the only term that violates CP in the Weyl anomaly and it will be the main focus of the present paper.

Since the Weyl transformation is Abelian, the Weyl anomaly must satisfy the simple Wess-Zumino consistency condition \cite{Bonora:1985cq}:
\begin{align}
0 = [\delta_{\sigma}, \delta_\tau] S= \int d^4x \left(\sigma\delta_\tau-\tau\delta_\sigma\right)\sqrt{g}T^m_m \ .
\end{align}
It is immediate to see that the both $\mathrm{Weyl}^2$ and $\mathrm{Pontryagin}$ terms satisfy the condition because they are Weyl invariant themselves.  On the other hand, $\Box R $ term is consistent but trivial because 
\begin{align}
\delta_\sigma\int d^4x\sqrt{g}R^2&=12\int d^4x\sqrt{g}\sigma \Box R \ 
\end{align}
and one can always remove it by adding the local counter term $R^2$. This also shows that  $R^2$ term (alone) in the Weyl anomaly does not satisfy the Wess-Zumino consistency condition.

The consistency of the Euler density is more non-trivial. For our purpose, let us introduce the Fradkin-Tseytlin-Riegert-Paneitz \cite{Fradkin:1982xc}\cite{Fradkin:1981jc}\cite{Riegert:1984kt}\cite{PA} operator
\begin{align}
 \triangle_4 =\nabla^m\left(\nabla_m\nabla_n+2R_{mn}-\frac{2}{3}Rg_{mn}\right)\nabla^n \ , 
\end{align}
which is a Weyl invariant generalization of the Laplacian squared in four-dimensions. We may now use the identity
\begin{align}
\delta_\sigma\sqrt{g}\left(\mathrm{Euler} -\frac{2}{3}\Box R\right) = -4\sqrt{g}\Delta_4\sigma 
\end{align}
to show that the Euler density satisfies the Wess-Zumino consistency condition after the integration by part.

The Weyl anomaly coefficients $c$ and $a$ are related to non-local terms of two-point function and three-point function of the energy-momentum tensor of a conformal field theory \cite{Osborn:1993cr} in the flat space-time. On the other hand, the detailed analysis of the conformal Ward identity tells that the non-local two-point and three-point functions of the energy-momentum tensor in four-dimensions do not contain terms violating CP \cite{Costa:2011mg}\cite{Zhiboedov:2012bm} in the flat space-time. This means that one cannot measure the Pontryagin density in the Weyl anomaly from the non-local correlation functions of the energy-momentum tensor. 

In contrast, the effect of the Pontryagin density in the Weyl anomaly is, if any, solely contained in the semi-local or local terms of the correlation functions of the energy-momentum tensor in the flat space-time. The semi-local terms are the correlation functions that include at least one coordinate space delta function. Such anomalies that are not supported by non-local conformal correlation functions are called ``impossible anomaly" in \cite{Nakayama:2018dig}. The Pontryagin density in the Weyl anomaly is such an example. We will see that the supersymmetric partner of the Pontryagin density in the Weyl anomaly also generates impossible anomalies.

\subsection{Seeley-DeWitt coefficient}

In free field theories, one may perform the path integral explicitly to compute the effective action and one may associate the Weyl anomaly with the Seeley-DeWitt coefficients \cite{DeWitt:1965jb}\cite{Gilkey:1995mj}\cite{Vassilevich:2003xt}, which we would like to review. Let us consider an elliptic operator $D$ acting on a collection of fields $\phi(x)$ (of a certain bundle).  Using the complete basis of the eigenvalue equation $D\phi_\lambda(x)=\lambda\phi_\lambda(x)$, we may define the heat kernel $K(s;x,y|D)$ as 
\begin{equation}
\begin{split}
K(s;x,y|D) = \left\langle x\right|e^{-sD}\left|y\right\rangle=\sum\phi^\dag_\lambda(x) e^{-s\lambda}\phi_\lambda(y) \ . 
\end{split}
\end{equation}
It is called the heat kernel because it satisfies the heat equation
\begin{align}
\left( \frac{\partial}{\partial s} + D \right)K(s;x,y|D)  = 0  
\end{align}
with the initial condition $K(s=0;x,y|D) = \delta(x-y) \mathbf{1} $.

The trace of this heat kernel has an asymptotic expansion as  
\begin{align}
\mathrm{Tr}\left(fe^{-sD}\right)&=\int d^dx\sqrt{g}\left\langle x\right|\mathrm{tr} f(x) e^{-sD}\left|x\right\rangle \cr
&\simeq\sum_{n\in\mathbb{N}}s^{\frac{n-d}{2}}a_n(f,D) \ . 
\end{align}
 The coefficient $a_n$  is called the Seeley-DeWitt coefficient of the heat kernel associated with the elliptic operator $D$. For most of our applications, $f(x)$ is just a function and we often use the local expression $b_n$ defined by 
\begin{equation}
\begin{split}
a_n(f,D)=\int d^d x f(x) \sqrt{g}b_n(D).
\end{split}
\end{equation}

There is a general recipe to compute the Seeley-DeWitt coefficient. Suppose that the elliptic operator $D$ is expressed as $D=-\nabla^m \nabla_m-E$, where $\nabla_m$ is a suitable covariant derivative acting on a section of $\phi$. In such cases it is straightforward to compute the lower-order Seeley-DeWitt coefficients. In particular, $b_4(D)$ can be explicitly computed as 
\begin{equation}
\begin{split}
\label{b4}
b_4(D)=\frac{1}{(4\pi)^2 360}\mathrm{tr}&\left(60\Box E+60RE+180E^2+12\Box R+5R^2 \right. \cr  
 &  \left. -2R^{mn}R_{mn} +2R^{mnrs}R_{mnrs} +30\Omega^{mn} \Omega_{mn} \right)
\end{split}
\end{equation}
where $\Omega_{mn}=[\nabla_m,\nabla_n]$ is the curvature two-form.

For example, let us consider a conformally coupled complex scalar $\phi$ with a $U(1)$ charge $q$. Here $D = -\nabla^m \nabla_m - \frac{1}{6}R$ and the covariant derivative includes the $U(1)$ gauge connection $A_m$: $\nabla_m \phi = \partial_m \phi + iq A_m \phi$. The above formula gives 
\begin{align}
b_4\left(-\nabla^2-\frac{1}{6}R\right)=\frac{1}{(4\pi)^2 360}\left(-4\Box R+6 \mathrm{Weyl}^2 -2 \mathrm{Euler} -60q^2F^{mn}F_{mn} \right) ,
\end{align}
where $F_{mn} = \partial_m A_n - \partial_n A_m$.  Note that a bosonic field contributes to the Weyl anomaly as $T^{m}_m= b_4(D)$ from the path integral representation of the effective action with the zeta function regularization while a fermionic field contributes to the Weyl anomaly with $T^m_m = -b_4(D)$.

Let us now consider a Euclidean Weyl fermion $\psi_{\alpha}$ in the $(1/2,0)$ representation of the Euclidean rotation group $SO(4)$, which we will call left-handed. The Euclidean action is given by
\begin{align}
S = \int d^4x \bar{\psi}_{\dot{\alpha}} \nabla^{\dot{\alpha} \alpha} \psi_\alpha \ ,
\end{align}
where the spinor covariant derivative $\nabla^{\dot{\alpha}\alpha} = (\bar{\sigma}^m)^{\dot{\alpha} \alpha} \nabla_m$ with respect to the spin connection $w_m^{\ a b}$ and the $U(1)$ gauge connection $A_m$ is given by 
\begin{align}
\nabla_m \psi_\alpha &= \partial_m \psi_{\alpha} +\frac{1}{2} w_m^{\ a b} (\sigma_{ab})_{\alpha}^{\ \beta} \psi_\beta  + iq A_m \psi_{\alpha} \ . 
\end{align}
In order to define a meaningful Euclidean action we also have to introduce an independent Weyl fermion $\bar{\psi}_{\dot{\alpha}}$ in the $(0,1/2)$ representation of $SO(4)$ (with the $U(1)$ gauge charge $-q$), which we will call right-handed. We stress that it is mandatory to introduce the fermions with the both chiralities to write down any sensible  action for a Euclidean Weyl fermion.

The classical equations of motion gives the left-handed Weyl equation
\begin{align}
\nabla^{\dot{\alpha} \alpha} \psi_{\alpha} = 0
\end{align}
as well as the right-handed Weyl equation
\begin{align}
\nabla^{\dot{\alpha} \alpha} \bar{\psi}_{\dot{\alpha}} = 0 \ .
\end{align}
While each of the Weyl equation is independently Euclidean invariant and not related with each other, the action principle demands the existence of the both simultaneously.

For the left-handed Euclidean Weyl fermion, the natural second-order differential operator to compute the Seeley-DeWitt coefficient is given by
\begin{equation}
\begin{split}
D^{\ \beta}_\alpha=\nabla_{\alpha\Dot{\alpha}}\nabla^{\Dot{\alpha}\beta} \ .
\end{split}
\end{equation}
This operator can be rewritten in  the standard form as
\begin{align}
D^{\ \beta}_{\alpha} &=\frac{1}{2}( (\sigma^m \bar{\sigma}^n + \sigma^n \bar{\sigma}^m)_{\alpha}^{\ \beta} + (\sigma^m \bar{\sigma}^n - \sigma^n \bar{\sigma}^m)_{\alpha}^{\ \beta} ) \nabla_m \nabla_n \\
&=-\nabla^m \nabla_m \delta^{\beta}_{\alpha} - \left(-\frac{R}{4}\delta^{\beta}_{\alpha} + i q F_{mn}(\sigma^{mn})^{\ \beta}_{\alpha} \right) .
\end{align}
To compute the Seeley-DeWitt coefficient, we also need the curvature two-form 
\begin{align}
(\Omega_{mn})^{\ \beta}_{\alpha} = \frac{1}{2}R_{mnab}(\sigma^{ab})^{\ \beta}_{\alpha}  + iq F_{mn} \delta^{\beta}_{\alpha} \ . 
\end{align}
Substituting these into the general formula \eqref{b4} and evaluating the spinor trace, we obtain 
\begin{align}
-b_4(\nabla_{\alpha\Dot{\alpha}}\nabla^{\Dot{\alpha}\beta})
=\frac{1}{(4\pi)^2 360}&\left(6\Box R+9\mathrm{Weyl}^2-\frac{11}{2}\mathrm{Euler}+\frac{15}{2}\mathrm{Pontryagin}  \right. \cr 
& \left. -120q^2F^{mn}F_{mn} - {180}q^2F^{mn}\Tilde{F}_{mn} \right), \label{b4c}
\end{align}
where $\tilde{F}_{mn} = \frac{1}{2}\epsilon_{mn rs}F^{rs}$ as usual. We have put the minus sign explicitly here to emphasize that it is a fermionic field.

Similarly, for the right-handed Euclidean Weyl fermion, the natural second-order differential operator is
\begin{equation}
\begin{split}
D^{\dot{\beta}}_{\ \dot{\alpha}}=\nabla^{\dot{\beta} \alpha}\nabla_{\alpha \dot{\alpha}}
\end{split}
\end{equation}
and the corresponding Seeley-DeWitt coefficient is
\begin{align}
-b_4(\nabla^{\dot{\beta} \alpha}\nabla_{\alpha \dot{\alpha}})
=\frac{1}{(4\pi)^2 360}&\left(6\Box R+9\mathrm{Weyl}^2-\frac{11}{2}\mathrm{Euler}-\frac{15}{2}\mathrm{Pontryagin}  \right. \cr 
& \left. -120q^2F^{mn}F_{mn} + {180}q^2F^{mn}\Tilde{F}_{mn} \right) . \label{b4a}
\end{align}
It is important to realize that the Seeley-DeWitt coefficient itself can be defined without introducing a fermion with the other chirality. 

As already observed in \cite{Christensen:1978md}\cite{Duff:1980qv}\cite{Nakayama:2012gu}, if we identified the Seeley-DeWitt coefficient of the Euclidean Weyl fermion of one chirality as the Weyl anomaly $T_{m}^m \stackrel{?}{=}-b_4(\nabla_{\alpha\Dot{\alpha}}\nabla^{\Dot{\alpha}\beta})$, we would obtain the Pontryagin density of the tangent bundle and the $U(1)$ gauge bundle. On the other hand, the sum of \eqref{b4c} and \eqref{b4a} does not contain the CP-violating terms. Note also that in the Lorentzian signature, the Pontryagin density gives an extra factor of $i$
\begin{align}
-b_4(\nabla_{\alpha\Dot{\alpha}}\nabla^{\Dot{\alpha}\beta})|_{\text{Lorentz}}
=\frac{1}{(4\pi)^2 360}&\left(6\Box R+9\mathrm{Weyl}^2-\frac{11}{2}\mathrm{Euler}+i\frac{15}{2}\mathrm{Pontryagin}  \right. \cr 
& \left. -120q^2F^{mn}F_{mn} -i {180}q^2F^{mn}\Tilde{F}_{mn} \right),
\end{align}
in the analytically continued Seeley-DeWitt coefficient. If we interpreted it as the Weyl anomaly of a physical theory, it would imply the violation of the unitarity.

\subsection{Effective action}
Since the Pontryagin density in the  Weyl anomaly satisfies the Wess-Zumino consistency condition, it should be integrable. Accordingly it should be possible to construct an effective action that reproduces the Pontryagin density in the Weyl anomaly as a classical variation. The question whether this is possible or not is distinct from the origin of the Pontryagin term in the Weyl anomaly and can be independently studied. 

At the minimal level, the effective action that gives the Pontryagin term in the Weyl anomaly turns out to be 
\begin{align}
S = \int d^4x \sqrt{g} \left( \frac{1}{2} \phi \Delta_4 \phi {+} Q\mathcal{Q}\phi - \phi \mathrm{Pontryagin}  \right) \ .
\end{align}
This classical action is also known as the dilaton effective action because it could appear as the Nambu-Goldstone action for the spontaneous breaking of the Weyl (or conformal) symmetry. 

Here, in addition to the Weyl invariant Fradkin-Tseytlin-Riegert-Paneitz operator $\triangle_4$, we introduced the so-called Q-curvature \cite{Branson}:
\begin{equation}
\begin{split}
 \mathcal{Q}&=-\frac{1}{6}\Box R-\frac{1}{2}R^{mn}R_{mn} +\frac{1}{6}R^2=\frac{1}{4}\left(\mathrm{Euler}-\mathrm{Weyl}^2 -\frac{2}{3}\Box R \right) ,
\end{split}
\end{equation}
which shows a remarkable property under the Weyl transformation
\begin{align}
\delta_{\sigma} \sqrt{g} \mathcal{Q} = - \sqrt{g} \triangle_4 \sigma \ . 
\end{align}
This results in the classical violation of the Weyl symmetry 
\begin{align}
T_m^m = + Q \triangle_4 \phi =-\frac{Q^2}{4} \left(\mathrm{Euler}-\mathrm{Weyl}^2 -\frac{2}{3}\Box R \right)  + Q \mathrm{Pontryagin} \ , 
\end{align}
where we have used the classical equation of motion for $\phi$. We see that the  classical variation contains the Pontryagin density as we claimed.

Alternatively, one may further assign the Weyl transformation of $\phi$ as $\phi \to \phi  +Q\sigma$. Under this compensated Weyl transformation,\footnote{Due to the linear shift under the Weyl transformation, the Weyl symmetry is spontaneously broken in this model.} the effective action changes as
\begin{align}
\delta S  =  \int d^4x \sqrt{g}\sigma(Q^2 \mathcal{Q} - Q \mathrm{Pontryagin}) \ ,
\end{align}
which directly gives the Weyl anomaly including the Pontryagin density.

One may compute the correlation functions of the energy-momentum tensor in the flat space by using this dilaton effective action. We first solve $\phi$ by using the equations of motion $\triangle_4\phi_c(x) =-Q\mathcal{Q}+ \mathrm{Pontryagin}$ from the Green function $\triangle_4 G_4(x-y) = \delta(x-y)$. Then we substitute it back into the action to evaluate the on-shell action. The three-point function of the energy-momentum tensor in a flat space can be computed as 
\begin{equation}
\begin{split}
\langle T_{mn} (x)T_{rs} (y)T_{ab} (z)\rangle&=\left.
\frac{(-2)^3}{\sqrt{g}\sqrt{g}\sqrt{g}}\frac{\delta}{\delta g^{mn}(x)}\frac{\delta}{\delta g^{rs}(y)}\frac{\delta}{\delta g^{ab}(z) }S[g,\phi=\phi_c]\right|_{g_{mn}=\delta_{mn}}.
\end{split}
\end{equation}
Since the result is lengthy, we only focus on the CP-violating part, which is our main focus. It is given by
\begin{equation}
\begin{split}
\langle T_{mn} (x)T_{rs} (y)T_{ab} (z)\rangle_{\mathrm{odd}}&=\frac{1}{3}Q\epsilon_{knlb}(\partial^k \partial_a \left[\partial^l \partial_m \delta(x-z) (\partial_r\partial_s-
\delta_{rs}\Box)G_2(x-y)\right]
\\&-\delta_{ma}\partial^k \partial^t\left[ \partial^l \partial_t \delta(x-z) (\partial_r \partial_s
-\delta_{rs }\Box)G_2(x-y)\right])+\mathrm{sym}.
\end{split}
\end{equation}
Here derivatives act on the $x$ coordinate and $G_2(x-y)$ is the Green function for the Euclidean Laplacian: $ \Box G_2(x-y) = \delta^{mn} \partial_m \partial_n G_2(x-y) = \delta(x-y)$. We see that the three-point function is only semi-local with one coordinate space delta function.

In particular, its trace gives  
\begin{equation}
\begin{split}
\langle T^{m}_m (x)T_{rs} (y)T_{ab} (z)\rangle_{\mathrm{odd}}&=-2Q\epsilon_{kslb}(\partial^z_r \partial^y_a-\delta_{ra}\partial^z_t {\partial^y}^{t}) \partial^l \delta(x-z) \partial^k \delta(x-y)+\mathrm{sym} \ ,
\end{split}
\end{equation}
which correctly reproduces the shape of the three-point function expected from the Pontryagin density in the Weyl anomaly.

A couple of comments are now in order. Firstly, we may introduce the additional term $ \phi \mathrm{Weyl}^2$ to generate the independent $\mathrm{Weyl}^2$ term in $T_m^m$. Secondly, as one can directly see, the three-point functions computed from the dilaton effective action is semi-local. It was under active debate if this could serve as the ``correct" Wess-Zumino action for the Weyl anomaly \cite{Deser:1993yx}\cite{Erdmenger:1996yc}\cite{Coriano:2017mux} because of the lack of the non-local contributions to the correlation functions.\footnote{In two-dimensions, the Liouville action or Polyakov action reproduces the local as well as non-local correlation functions of the energy-momentum tensor.} As for the Pontryagin density, however, there is no non-local term from the beginning, so the debate was irrelevant for us. Finally, the dilaton effective action can be modified in various manners by adding more Weyl invariant terms. They give differently looking effective actions for various purposes. For example, \cite{Komargodski:2011vj} studied a variant with the two-derivative kinetic term to prove the a-theorem while \cite{Levy:2018bdc} studied the quantum nature by adding the Liouville potential. 

\section{CP-violating super Weyl anomaly}
\subsection{General structure}

In order to study the supersymmetric extension of the Pontryagin density in the Weyl anomaly, we first review the structure of the super  Weyl anomaly. Let us consider the supercurrent multiplet with the supersymmetric conservation law
\begin{align}
\bar{\mathcal{D}}^{\dot{\alpha}} T_{\alpha \dot{\alpha}} + \frac{2}{3}\mathcal{D}_{\alpha} T = 0 \ ,
\end{align}
where $T$ is a chiral superfield known as the supertrace multiplet. See e.g. \cite{Komargodski:2010rb} for a detailed analysis of this equation in the flat space-time. 
For superconformal field theories, $T=0$ in the trivial supergravity background. In the non-trivial supergravity background, the supertrace multiplet shows the super Weyl anomaly \cite{Bonora:1984pn}:
\begin{align}
 8\pi^2 T &= c W^{\alpha\beta\gamma} W_{\alpha\beta\gamma} - a\left(W^{\alpha\beta\gamma} W_{\alpha\beta\gamma}  - \frac{1}{4}(\bar{\mathcal{D}}^2-4R) (G^mG_m + 2R\bar{R})\right) \cr
 &+ \frac{1}{16} h(\bar{\mathcal{D}}^2 - 4R) \mathcal{D}^2 R \ . \label{superW}
\end{align}
Here  $W^{\alpha\beta\gamma}$ is the chiral super Weyl tensor multiplet, and $W^{\alpha\beta\gamma} W_{\alpha\beta\gamma}  - \frac{1}{4} (\bar{\mathcal{D}}^2 -4R) (G^m G_m  + 2R\bar{R})$ is the chirally projected super Euler density (see e.g.  \cite{Buchbinder:1998qv}).
In the literature it is usually assumed that $a$ and $c$ are real, but we will relax the condition soon. The term proportional to the real parameter $h$ is  trivial and can be removed by using appropriate supersymmetric local counterterm.

Note that the chiral multiplet $T$ contains the trace of the energy-momentum tensor and the divergence of the R-current $-\frac{1}{8} \left(T^{m}_{m} + \frac{3}{2}i \nabla^m J_m^R\right)$ in the  $\theta^2$  component. See e.g. \cite{Komargodski:2010rb}. Thus, in component, assuming $a$ and $c$ are real for a moment, super Weyl anomaly \eqref{superW} implies the Weyl anomaly (up to trivial terms)
\begin{align}
T^{m}_{m} =  \frac{c}{16\pi^2} \mathrm{Weyl}^2 - \frac{a}{16\pi^2} \mathrm{Euler}- \frac{c}{6\pi^2}F^{mn}F_{mn}
\end{align}
as well as the chiral anomaly for the R-current conservation
\begin{align}
\nabla^m J_m^R|_{\mathrm{Lorentz}} = \frac{c-a}{24\pi^2} R^{mnrs}\tilde{R}_{mnrs} + \frac{5a-3c}{27\pi^2} F^{mn}\tilde{F}_{mn} \ .
\end{align}
Here $F_{mn}$ is the field strength for the R-symmetry gauge field (i.e. graviphoton field strength). See e.g. Appendix A of \cite{Anselmi:1997am} for the detailed derivation.\footnote{The formula is corrected in \cite{Cassani:2013dba}. We would like to thank I.~Papadimitriou for the correspondence.} 

Alternatively one may study the variation of the supersymmetric effective action  under the super Weyl variation $\mathcal{E} \to e^{3\sigma} \mathcal{E}$.  The anomalous variation shows
\begin{align}
\delta S &= \int d^4x d^2\theta \mathcal{E} \sigma T +   \int d^4x d^2\bar{\theta} \bar{\mathcal{E}} \bar{\sigma} \bar{T} \ , 
\end{align}
where $\sigma$ is a chiral superfield corresponding to the super Weyl variation. In component, $\sigma = \sigma_1 + i\sigma_2 + O(\theta)$ with $-\frac{\sigma_1}{2}$ being the Weyl factor\footnote{In our convention that follows \cite{Buchbinder:1998qv}, the sign of the  Weyl factor and the super Weyl factor is opposite.} and $\sigma_2$ being the gauge parameter for the R-symmetry transformation.

It is known that (up to terms that can be removed by local counterterms), these are the only available super Weyl anomaly (see e.g. \cite{Bonora:2013rta} and older reference therein). However, there is one fine print that has not been discussed very much in the literature, namely the reality condition on $a$ and $c$ (as well as $h$). One may try to imagine what happens if $a$ and $c$ are not real but complex numbers. 

First of all, the Wess-Zumino consistency condition demands that $a$ must be real \cite{Auzzi:2015yia} from the coefficients of
\begin{align}
(\mathcal{D}^\alpha \sigma \bar{\mathcal{D}}^{\dot{\alpha}} \bar{\tau} G_{\alpha \dot{\alpha} }) - (\sigma \leftrightarrow \tau) \ . 
\end{align}
One can also see how this must be the case from the component analysis. If $a$ were not real, then $\nabla ^{m} J_{m}^R$ would include the term proportional to the Euler density whose Weyl variation is non-zero. Then it could not satisfy the mixed Wess-Zumino consistency condition for the Weyl transformation and the (gauged) R-symmetry transformation. Similarly one can show that $h$ must be real.

However, there is no reality constraint on $c$ from the Wess-Zumino consistency condition simply because $W^{\alpha\beta\gamma}W_{\alpha\beta\gamma}$ is super Weyl invariant. Assuming $c$ is a complex number, we have the CP-violating Weyl anomaly
\begin{align}
T^{m}_{m}|_{\text{Lorentz}} =& \frac{\mathrm{Re}(c)}{16\pi^2} \mathrm{Weyl}^2 - \frac{a}{16\pi^2} \mathrm{Euler}- \frac{\mathrm{Re}(c)}{6\pi^2}F^{mn}F_{mn}  \cr
 &- \frac{\mathrm{Im}(c)}{16\pi^2} \mathrm{Pontryagin} +  \frac{\mathrm{Im}(c)}{6\pi^2}F^{mn}\tilde{F}_{mn} 
\end{align}
and the corresponding parity-violating chiral anomaly for the R-current
\begin{align}
\nabla^m J_m^R|_{\text{Lorentz}} =&  \frac{\mathrm{Re}(c)-a}{24\pi^2} R^{mnrs}\tilde{R}_{mnrs} + \frac{5a-3\mathrm{Re}(c)}{27\pi^2} F^{mn}\tilde{F}_{mn}  \cr
&+ \frac{\mathrm{Im}(c)}{24\pi^2} \mathrm{Weyl}^2 - \frac{\mathrm{Im}(c)}{9\pi^2} F^{mn}F_{mn}  \ .
\end{align}
 We can check that these satisfy the mixed Wess-Zumino consistency condition of the Weyl transformation and the R-symmetry transformation. 

One thing to be noted here is that these two expressions are written in the Lorentzian signature and they are all real. As a consequence they are compatible with unitarity. In the  Euclidean signature we have to put $i$ in front of the $\epsilon$ tensor so that the supersymmetric Pontryagin term in the Weyl anomaly gives a pure phase in the Euclidean partition function. Similarly, the CP-violating anomaly in the R-symmetry transformation is not a phase but an absolute value in the Euclidean partition function.

Note that it is not feasible to absorb the phase of $c$ into the definition of the chiral superfield $\sigma$ that defines the super Weyl transformation. This is because $a$ must be real and such a redefinition would violate the Wess-Zumino consistency condition. In this way, having a imaginary part of $c$ leads to a physically non-trivial effect.

Let us take a closer look at the structure of the supersymmetric partners of the Pontryagin density in the Weyl anomaly. We note that these novel anomalies break parity or at least CP. The Weyl anomaly contains a new term of the graviphoton $\theta$ term $ F^{mn}\tilde{F}_{mn}$ associated with the R-symmetry gauging. Its existence would be related to the renormalization group beta functions for the graviphoton $\theta$ term. The R-current anomaly now contains unfamiliar CP-violating terms. One is the graviphoton field strength squared and the other is the Weyl tensor squared.

Not only they look unfamiliar, but also all of them are examples of impossible anomalies. Non-local terms in the three-point functions among the energy-momentum tensor and R-current do not contain any terms that will directly generate these CP-violating anomalies. Instead all of them are supported in the semi-local terms. For example, the R-current anomaly has the CP-violating semi-local terms proportional to
\begin{align}
\langle J^R_\mu(x) J^R_\nu(y) J^R_\rho(z) \rangle = \partial^y_{\alpha} (\partial_\mu G_2(x-y) \delta_{\nu\rho} \partial^{\alpha}\delta(y-z)) - \partial^y_\nu (\partial_\mu G_2(x-y) \partial_\rho \delta(y-z)) + \mathrm{sym} \   
\end{align}
in a particular regularization scheme.

\subsection{Supersymmetric Seeley-DeWitt coefficient}
Similarly to what we have studied in section 2.3, in free supersymmetric field theories, the super Weyl anomaly is related to the supersymmetric Seeley-DeWitt coefficients. Here we would like to study a supersymmetric generalization of the heat kernel computation. 

Let us first define the supersymmetric heat kernel for a chiral operator on the superspace
\begin{align}
\frac{\partial U_c}{\partial s} = \left(\Box_+ -\frac{1}{4}(\bar{\mathcal{D}}^2 \bar{R}) + R\bar{R} \right) U_c \ 
\end{align}
with the initial condition 
\begin{align}
U_c(s=0;z,z') = \delta_+(z,z') \ . 
\end{align}
Here $\delta_+(z,z')$ is a covariantly chiral delta function, and the chiral Laplacian $\Box_+$ is defined by
\begin{align}
\Box_+ = \mathcal{D}^m \mathcal{D}_m + \frac{1}{4} R \mathcal{D}^2 + iG^m \mathcal{D}_m + \frac{1}{4}(\mathcal{D}^\alpha R)\mathcal{D}_{\alpha} \ .
\end{align}

The supersymmetric Seeley-DeWitt coefficient $a^c_n$ for a chiral superfield is defined as an asymptotic power-series solution of its trace:
\begin{align}
U_c(s;z,z) = \frac{1}{(4\pi s)^2} \sum_{n=0}^{\infty} a^c_n s^n  \ .  
\end{align}
The most relevant one for our discussion is $a_2^c$ and it is given by \cite{McArthur:1983vk}\cite{McArthur:1983fk}\cite{Buchbinder:1986im}
\begin{align}
a_2^c = \frac{1}{12} W^{\alpha\beta\gamma} W_{\alpha\beta\gamma} + \frac{1}{48}(\bar{\mathcal{D}}^2 -4R) G^m G_m  -\frac{1}{96}(\bar{\mathcal{D}}^2-4R)(\mathcal{D}^2-4\bar{R})R \ . 
\end{align}

Similarly, if we started with the anti-chiral operator $\Box_{-}$, we would end up with the supersymmetric Seeley-DeWitt coefficient $\bar{a}^a_n$ for an anti-chiral superfield. In particular, 
\begin{align}
\bar{a}_2^a = \frac{1}{12} \bar{W}^{\alpha\beta\gamma} \bar{W}_{\alpha\beta\gamma} + \frac{1}{48}({\mathcal{D}}^2 -4\bar{R}) G^m G_m  -\frac{1}{96}({\mathcal{D}}^2-4\bar{R})(\bar{\mathcal{D}}^2-4{R}) \bar{R} \ . 
\end{align}

Suppose we have a supersymmetric massless Wess-Zumino model coupled with background superconformal supergravity. It is described by the classical action
\begin{align}
S[\Phi] = \int d^4x d^2 \theta d^2\bar{\theta} E^{-1} \bar{\Phi} \Phi \ ,
\end{align}
where $\Phi$ and $\bar{\Phi}$ are chiral and anti-chiral superfield. One can compute the effective action explicitly with the supersymmetric zeta function regularization of the path integral, and then one can relate the super Weyl anomaly under the super Weyl transformation $\mathcal{E} \to e^{3\sigma} \mathcal{E}$ with the supersymmetric Seeley-DeWitt coefficient:
\begin{align}
\delta S = \frac{1}{(4\pi)^2}\left(\int d^4x  d^2\theta \mathcal{E}  \sigma a_2^c + \int d^4x d^2\bar{\theta} \bar{\mathcal{E}} \bar{\sigma} \bar{a}_2^a  \right)\ .
\end{align}
This is the standard expression for the super Weyl anomaly of a free chiral multiplet (i.e. one complex scalar and one Weyl fermion) with the identification $T = \frac{1}{(4\pi)^2}a_2^c$.  In particular, $a = \frac{1}{48}$ and $c = \frac{1}{24}$ are real numbers.  Consequently it does not show the Pontryagin density in the Weyl anomaly.\footnote{In order to compare it with the component expression of the Weyl anomaly, we should note that superconformal R-charge $q$ of a conformal scalar is $2/3$ and that of a Weyl fermion is $-1/3$.}

What would be the corresponding supersymemtric Seeley-DeWitt coefficient for a Euclidean left-handed Weyl fermion (without a right-handed partner)? The idea is that in the Euclidean signature, one may take the super Weyl parameter $\sigma$ and $\bar{\sigma}$ independently, In particular we can set $\bar{\sigma} = 0$ (while $\sigma$ is nonzero). This corresponds to considering the Weyl transformation on $\Phi$ (that contains the left-handed Weyl fermion) without the Weyl transformation on $\bar{\Phi}$ (that contains the right-handed Weyl fermion). 

The resultant super  Weyl variation is
\begin{align}
\int d^4x d^2\theta  \mathcal{E} \sigma \left( \frac{1}{12} W^{\alpha\beta\gamma} W_{\alpha\beta\gamma} + \frac{1}{48}(\bar{\mathcal{D}}^2 -4R) G^m G_m  -\frac{1}{96}(\bar{\mathcal{D}}^2-4R)(\mathcal{D}^2-4\bar{R})R \right) \ 
\end{align}
without the anti-chiral part (i.e. the terms with $\bar{\sigma}$). Suppose we would like to study the response to the Weyl transformation of this theory from the coupling to the real part of $\sigma$.  Then, in components, we see that the would-be Weyl anomaly (which is given by the coupling to the real part of $\sigma$) includes the Pontryagin density with a real coefficient in the Euclidean signature.\footnote{After continuing to the Lorentzian signature, this would give an imaginary coefficient in front of the Pontryagin density in the would-be Weyl anomaly, which is consistent with what we saw in section 2. This is to be contrasted with the manifestly real coefficient of the  Pontryagin density in the Weyl anomaly discussed in section 3.1.}

\subsection{Supersymmetric effective action}
As we have just seen in the previous subsection, there is no simple free field computation that will give the supersymmetric generalization of the Pontryagin density in the Weyl anomaly. However, one can still construct the supersymmetric dilaton effective action to incorporate the Pontryagin density in the super Weyl anomaly.

For this purpose, we would like to generalize the $\mathcal{N}=1$ super Liouville theory studied in \cite{Levy:2018xpu} (see also \cite{Butter:2013ura}\cite{Bobev:2013vta}). The classical action for a chiral superfield $\Phi$ is given by
\begin{align}
S[\Phi] = \int d^4x d^2 \theta \mathcal{E} \left(\Phi \hat{\mathcal{P}} \bar{\Phi} + 4 \bar{Q} \hat{\mathcal{Q}} \Phi  + C\Phi W^{\alpha\beta\gamma}W_{\alpha\beta\gamma}\right) + h.c. \ .
\end{align}
Here the chirally projected supersymmetric Fradkin-Tseytlin-Riegert-Paneitz operator $\hat{\mathcal{P}}$ \cite{Butter:2013ura} is given by 
\begin{align}
\hat{\mathcal{P}} = -\frac{1}{64}\left(\bar{\mathcal{D}}^2 - 4R\right)(\mathcal{D}^2\bar{\mathcal{D}}^2 + 8 \mathcal{D}^\alpha (G_{\alpha\dot{\alpha}}\bar{\mathcal{D}}^{\dot{\alpha}})) \ ,
\end{align}
and the supersymmetric Q-curvature chiral superfield \cite{Butter:2013ura}\cite{Levy:2018xpu} is given by 
\begin{align}
\hat{\mathcal{Q}} = -\frac{1}{8}(\bar{\mathcal{D}}^2-4R)\left(G^mG_m + 2 R\bar{R} - \frac{1}{16} \mathcal{D}^2 R \right) \ .
\end{align}

Note that in \cite{Levy:2018xpu}, it is assumed that $Q$ is a real parameter, but here we would like to regard it as a complex parameter for the most genericity. We also note that the parameter $C$ is complex. Under the super Weyl variation with additional shift of $\Phi$
\begin{align}
\Phi &\to \Phi - 2Q\sigma \cr
\bar{\Phi} &\to \bar{\Phi} - 2\bar{Q} \bar{\sigma} \ , 
\end{align}
the supersymmetric dilaton effective action transforms as 
\begin{align}
S[\Phi] \to S[\Phi] - S[2Q\sigma] \ .
\end{align}
The infinitesimal variation is
\begin{align}
\delta S[\Phi] = -\int d^4x d^2 \theta \mathcal{E} \sigma \left(8 \bar{Q}Q \hat{\mathcal{Q}}  + 2QC W^{\alpha\beta\gamma}W_{\alpha\beta\gamma}  \right) + h.c. 
\end{align}
which gives the supe Weyl variation studied e.g. in \cite{Bonora:2013rta}\cite{Auzzi:2015yia} except that in our case $QC$ can be a complex number, whose possibility has not been emphasized before. Note also that the coefficient in front of $\hat{\mathcal{Q}}$ is a real number, whose necessity is not immediately obvious but the Wess-Zumino consistency condition of the super Weyl anomaly demands it must be the case \cite{Auzzi:2015yia}. It is a non-trivial check that our supersymmetric dilaton effective action with a complex $Q$ consistently  generates the real coefficient here.

We now discuss the component form of the super Weyl anomaly. The real component of the chiral superfield $\sigma$ is the usual Weyl variation while the pure imaginary component of $\sigma$ is the gauge parameter for the R-symmetry. Expressing the super Weyl anomaly in the conventional form
\begin{align}
{T} = \frac{c}{8\pi^2}W^{\alpha\beta\gamma} W_{\alpha\beta\gamma} - \frac{a}{8\pi^2} \left( W^{\alpha\beta\gamma} W_{\alpha\beta\gamma} - \frac{1}{4}(\bar{\mathcal{D}}^2 - 4R)(G^m G_m+2R\bar{R}) \right) \  
\end{align}
together with the anti-chiral part $\bar{T}$,
we have the (classical) identification
\begin{align}
c &= 32\pi^2 \bar{Q}Q - 16\pi^2 QC \cr
a & = 32\pi^2\bar{Q}Q \ .
\end{align}
When $QC$ is not a real number, we have the classical Pontryagin density in the Weyl anomaly from the imaginary part of $QC$ as the variation of the supersymmetric dilaton effective action. We note that $a$ remains real even for a complex $Q$ as is expected from the Wess-Zumino consistency condition.

\section{Conclusion}

In this paper, we have discussed the supersymmetric completion of the CP-violating Pontryagin density in the Weyl anomaly. For this purpose, it was crucial to complexify the central charge $c$, where the existence of the imaginary part leads to the Pontryagin density in the Weyl anomaly as well as other CP-violating terms in the supersymmetric Weyl anomaly.

The consistency of the complexified $c$ can be seen from the perspective of the string theory as well. In the Calabi-Yau compactification of the type II string theory, the effective coupling constant for the gravitational $F$-term (i.e. the Weyl tensor squared and the  Pontryagin density \cite{Deser:1980kc}) can be obtained from
\begin{align}
S = \int d^4x d^2\theta\mathcal{E} F_1(t) W^{\alpha\beta\gamma} W_{\alpha\beta\gamma} + h.c.  \ ,
\end{align} 
where $F_1(t)$ is the genus one topological string amplitude \cite{Antoniadis:1992sa}\cite{Antoniadis:1992rq}. Note that the gravitational coupling $F_1(t)$ is a ``holomorphic" function of the moduli fields $t$. Its real part determines the coupling constant for the Weyl tensor squared and its imaginary part determines that for the Pontryagin density. 

The supersymmetric Weyl anomaly is nothing but the renormalization group beta function for $F_1(t)$, and since $F_1(t)$ is holomorphic it should be consistent to complexify the beta function as well. To be more precise, $F_1(t)$ has a holomorphic anomaly that reflects the non-locality of the effective action \cite{Bershadsky:1993cx}. The holomorphic anomaly equation that determines the non-holomorphicity is closely related to the supersymmetric Weyl anomaly although there is an extra stringy contribution to it. It would be an interesting future direction  to see why we obtain a ``real" beta function for the complex coupling constant in most examples from the viewpoint of the string theory. Furthermore, we would like to pursue if there is any chance to obtain the imaginary part to induce the Pontryagin density in the super Weyl anomaly in the string setup.

We should emphasize that the Pontryagin density obtained in this way is a real number in the Lorentzian signature and does not violate the unitarity. In section 3.2, we have addressed the other possibility that the super Weyl variation of the effective action can be only holomorphic with respect to the super Weyl parameter $\sigma$. This is feasible in the Euclidean signature, and it would lead to the Pontryagin density with an imaginary coefficient (if we naively analytically continue to the Lorentzian signature at the sacrifice of unitarity) under the holomorphic (i.e. chiral) super Weyl variation. 

We have the analogous situation in stringy inspired theories of ``non-anticommutative" field theories, or $\mathcal{N}=1/2$ supersymmetric field theories \cite{Ooguri:2003tt}\cite{Seiberg:2003yz}. They may arise from the self-dual graviphoton condensate in the Euclidean string theory. There, the available classical symmetry is the holomorphic super Weyl variation of $\mathcal{E} \to e^{3\sigma} \mathcal{E}$ without the anti-holomorphic partner. The absence of the anti-holomorphic variation (i.e. $\bar{\sigma})$ is because the theory has no supersymmetry for the $\bar{\theta}$ translation, giving the name of $\mathcal{N}=1/2$ supersymmetry. Then the natural ``supersymmetric Weyl anomaly" for the $\mathcal{N}=1/2$ supersymmetric field theories would be similar to the supersymmetric (chiral) Seeley-DeWitt coefficient studied in section 3.2 and it could lead to the Pontryagin density with an imaginary coefficient under the holomorphic (or chiral) super Weyl variation. The background might be closely related to the axial-gravity studied in \cite{Bonora:2018obr} and it may be worthwhile studying its connection.

\section*{Acknowledgements}
One of the authors (Y.~N.) would like to thank Keren-Zur Boaz, Sergei Kuzenko and Yaron Oz for the correspondence. 
This work is in part supported by JSPS KAKENHI Grant Number 17K14301.

\appendix

%%%%%%%%%%%%%%%%%%%%%%%%%%%%%%%%%%%%%%%%%%%%%%%%%%%%%%%%%%%%%%%%%%%%%%%%%%%%%%%%%%

\end{document}